\documentclass[copyright,creativecommons]{eptcs}
\providecommand{\event}{DCM'09}
\providecommand{\volume}{}
\providecommand{\anno}{2009}
\providecommand{\firstpage}{1}

\usepackage[dvips]{color}
\usepackage{graphicx}
\usepackage{amssymb}
\usepackage{multicol}
\usepackage{mathrsfs}
\usepackage{breakurl}

\def\grameq{$::=$}

\definecolor{rdxbackcolor}{gray}{0.91}
\def\rdx#1{\smash{\colorbox{rdxbackcolor}{\strut\smash{$#1$}}}}
\newcommand{\substitute}[2]{_{\{#1/#2\}}}

\newtheorem{theorem}{Theorem}
\newtheorem{definition}[theorem]{Definition}
\newtheorem{example}[theorem]{Example}

\title{An Intuitive Automated Modelling Interface\\ for Systems Biology}
\author{Ozan Kahramano\u gullar{\i} 
\institute{
The Microsoft Research -- University of Trento\\ 
Centre for Computational and Systems Biology
\thanks{This work has been done during Kahramano\u gullar{\i}'s appointment 
at the Department of Computing, Imperial College and 
Centre for Integrative Systems Biology at Imperial College.
Kahramano\u gullar\i \space acknowledges support of the UK Biotechnology and Biological 
Sciences Research Council through the Centre for Integrative Systems Biology at Imperial College 
(grant BB/C519670/1).}
}
\and
Luca Cardelli
\institute{Microsoft Research Cambridge\\
Centre for Integrative Systems Biology at Imperial College}
\and \\
 Emmanuelle Caron 
\institute{
Centre for Integrative Systems Biology at Imperial College\\ 
Centre for Molecular Microbiology and Infection, Imperial College}
}
\def\titlerunning{An Intuitive Automated Modelling Interface for Systems Biology}
\def\authorrunning{{Kahramano\u gullar{\i}}, Cardelli  \& Caron}
\begin{document}
\maketitle

\begin{abstract}
We introduce a natural language interface for building stochastic
$\pi$ calculus models of biological systems. In this language, complex
constructs describing biochemical events are built from basic
primitives of association, dissociation and transformation. This language thus allows us to  model biochemical systems modularly by describing their dynamics in a narrative-style language, while making amendments, refinements and extensions on the models easy.  We demonstrate the language on a model of Fc$\gamma$ receptor phosphorylation during phagocytosis. We provide a tool implementation of the translation into a stochastic $\pi$ calculus language, Microsoft Research's \textsf{SPiM}.
\footnote{We dedicate this paper to the memory of Dr. Emmanuelle Caron, 
who unexpectedly passed away in July 2009.  It has been an honour to have worked with Emmanuelle, a biologist of the highest calibre.}
\footnote{This work has been presented as oral presentation at the BioSysBio'09 Conference and Noise in Life'09 Meeting, both held in Cambridge in March 2009.}
 \end{abstract}


\section{Introduction}

Modelling of biological systems by mathematical and computational techniques is becoming increasingly 
widespread in research on biological systems. 
In recent years, pioneered by Regev and Shapiro's seminal 
work \cite{PRSS01,RS02}, there has been a considerable 
amount of research on applying computer science technologies
to modelling biological systems. Along these lines,
various languages with stochastic simulation 
capabilities  based on, for example, 
process algebra, term rewriting (see, e.g, \cite{DFFHK07,DFFK08})
and Petri nets (see, e.g., \cite{TTKLL07,HGD08}) have been proposed.  
However, expressing biological knowledge in specialised 
modelling languages often requires a simultaneous understanding of the  biological system
and expert knowledge of the modelling language. Isolating and communicating the biological knowledge
to build models for simulation and analysis is a challenging task both for wet-lab  biologists and modellers. 
 
Writing programs in simulation languages requires specialised training, and it is difficult even for the experts
 when complex interactions between biochemical species in biological systems are considered: the 
 representation of different states of a biochemical species with respect to all its interaction capabilities results 
 in an exponential blow up in the number of states. For example, when a protein with $n$ different interaction 
 sites is being modelled, this results in $2^n$ states, which needs to be represented in the model. Enumerating 
 all these states by hand, also without inserting typos is a difficult task.

To this end, we introduce an intuitive front-end interface language for building 
process algebra models of biological systems:  
 process algebras are languages that have originally been designed to formally describe complex reactive 
computer systems. Due to the resemblance between these computer systems and biological systems, process
 algebra have been recently used to model biological systems. An important feature of the process algebra 
 languages is the possibility to describe the components of a system separately and observe the emergent 
 behaviour from the interactions of the components (see, e.g., \cite{CCGKP08b,CCGKP08}).  
 
 Our  focus here is on
 $\pi$ calculus \cite{Mil99}, which is a broadly studied process algebra,
  also because of its compactness,  generality, and flexibility. 
Because  biological systems are typically highly complex 
and massively parallel, the $\pi$ calculus is well 
suited to describe their dynamics. In particular,
it allows the components of a biological system to be 
modelled independently, rather than modelling individual 
reactions. This allows large models to be constructed by 
composition of simple components. $\pi$ calculus also enjoys an 
expressive power in the setting of biological models that exceeds, 
e.g., Petri nets \cite{CZ08b}.

In the following, we present a language that consists of basic primitives of 
association, dissociation and transformation. We  impose certain consistency 
constraints on these primitive expressions, which are required  for the models 
that describe the dynamics of biochemical processes.  
We give a translation algorithm into stochastic $\pi$ calculus.  Based on this, 
we present the implementation of  a tool for 
automated translation of models into Microsoft Research's stochastic simulation language $\mathsf{SPiM}$ \cite{PCC06,PL07}, which can be used  
to run stochastic simulations on $\pi$ calculus models.  We demonstrate the language on a model 
of Fc$\gamma$ receptor phosphorylation during phagocytosis. 
 The implementation 
of the translation tool as well as further information is available for download at our 
website~\footnote{\texttt{http://www.doc.ic.ac.uk/$\sim$ozank/pim.html}}.

\section{Species, Sites, Sentences and Models}
\label{section:data:structure}
We follow the abstraction of biochemical species 
as stateful entities with connectivity interfaces to other species: a species can have a number of sites  
through which it interacts with other species,  and changes its state as a result of the interactions 
\cite{DFFHK08,GHP07}. In Section \ref{section:language}, we use this idea to design a natural language-like
  syntax for building models. The models written in this language are then automatically 
translated into a $\mathsf{SPiM}$ program by a translation algorithm. This is done by mapping 
the sentences of the language into events constructed from the basic primitives, 
which are then compiled into executable process expressions in the $\mathsf{SPiM}$ language.

There is a countable set of species $A,B,C,\ldots$.
Each species has a number of sites $a,b,c,\ldots$ with which it can bind to other 
species or unbind from other species when they are already bound.  
We write sentences that describe  the `behaviour' of each species with respect to their sites.  
There are three kinds of sentences, i.e., \emph{associations}, \emph{dissociations}, and \emph{transformations}.  
We define the sentences as 
$$
\langle \, \mathsf{type}, (A,a),\, (B,b), \, \textit{Pos}, \, \textit{Neg},\, r \, \rangle
$$
where $\mathsf{type} \in \{$\textit{association}, \textit{dissociation}, \textit{transformation}$\}$ 
is the type of the sentence. 
The pairs $(A,a)$ and $(B,b)$ are called the \emph{body}
of the sentence. The sets  \textit{Pos} and \textit{Neg} are called the \emph{conditions} of the sentences. 
$(A,a)$ and $(B,b)$ are pairs of species and sites, and \textit{Pos} and \textit{Neg}
are sets of such pairs of species and sites. If the sentence is an \emph{association}, it describes 
the event where the site $a$ on species $A$ associates to  the site $b$ on species $B$ 
if the sites on species in \textit{Pos} are already bound and those in \textit{Neg} are already unbound.
If it is a \emph{dissociation} sentence, it describes the dissociation of   
the site $a$ on species $A$ from the site $b$ on species $B$. 
A \emph{transformation} sentence describes the event of species $A$ transforming into species $B$,
where $B$ can be empty. In this case, this describes the decay of species $A$.
In transformation sentences, sites $a$ and $b$ must be  empty, since transformations are 
site independent. $r \in \mathbb{R}^{+}$ denotes the rate of the event 
that the sentence describes.  Then a model $\mathcal{M}$ is a set of such sentences. 
In Section \ref{section:language}, we give a representation of  these sentences in natural-language. 
For example, a sentence of the form  
$\langle \, \textit{association}, (A,a),\, (B,b), \, \{ (A,c)\}, \, \{\,\},\, 1.0 \, \rangle$
is given with the following English sentence. 
$$
\texttt{site a on A associates site b on B with rate 1.0 if site c on A is bound}
$$
We denote with $\mathsf{species}(\mathcal{M})$ all the 
species occurring in the  body of the sentences of $\mathcal{M}$. 
The function $\mathsf{sites}(\mathcal{M},A)$ denotes the sites of the species $A$ that
occur in the body of all the sentences of $\mathcal{M}$.
$\mathsf{sites}(\textit{Pos},A)$ denotes the sites of the species $A$ in $\textit{Pos}$ 
(similarly for $\textit{Neg}$). For any set $\mathcal{A}$, 
$\wp(\mathcal{A})$ denotes the powerset of $\mathcal{A}$.
 
\subsection{Conditions on Sentences} \label{subsection:conditions}
Given a model $\mathcal{M}$, we impose several conditions on its sentences.
 
\begin{enumerate}
\item
\textbf{Sentences contain relevant species.}  
The species in the condition of each sentence must 
be a subset of those in the body of the sentence.

\item
\textbf{Conditions of the sentences are consistent.}  
For every sentence of the form\\ 
$\langle\, \mathsf{type}, (A,a),\, (B,b), \, \textit{Pos}, \, \textit{Neg},\, r \, \rangle$,
we have that $\textit{Pos} \cap \textit{Neg} = \emptyset$.

\item
\textbf{All the sites in the conditions are declared in the model.}  
For every sentence of the form 
$\langle\, \mathsf{type}, (A,a),\, (B,b), \, \textit{Pos}, \, \textit{Neg},\, r \, \rangle$,
we have that 
$\mathsf{sites}(\textit{Pos},A) \subseteq \mathsf{sites}(\mathcal{M},A)$, \\ 
$\mathsf{sites}(\textit{Neg},A) \subseteq \mathsf{sites}(\mathcal{M},A)$,
$\mathsf{sites}(\textit{Pos},B) \subseteq \mathsf{sites}(\mathcal{M},B)$ and  
$\mathsf{sites}(\textit{Neg},B) \subseteq \mathsf{sites}(\mathcal{M},B)$.

\item
\textbf{Association sentences associate unbound species.}
For every association sentence \\
$\langle\, \textit{association}, (A,a),\, (B,b), \, \textit{Pos}, \, \textit{Neg},\, r \, \rangle$, we have that 
 $(A,a)$, $(B,b) \in \textit{Neg}$.

\item
\textbf{Dissociation sentences dissociate bound species.}
For every dissociation sentence \\
$\langle\, \textit{dissociation}, (A,a),\, (B,b), \, \textit{Pos}, \, \textit{Neg},\, r \, \rangle$, we have that 
 $(A,a)$, $(B,b) \in \textit{Pos}$. 

\item
\textbf{Transformation sentences are unbound at all sites.}
For every transformation sentence \\
$\langle\, \textit{transformation}, A,\, B, \, \textit{Pos}, \, \textit{Neg},\, r \, \rangle$, 
we have that 
$\textit{Pos} = \emptyset$ and $\textit{Neg} = \{ (A,x) \, | \, x \in \mathsf{sites}(\mathcal{M},A)\}$.
\end{enumerate}

When these conditions hold, we can map the sentences of a model to another representation
where the role of the conditions become more explicit.  
In the following, for a model $\mathcal{M}$,  we describe the states of its species as subsets of 
its sites, where bound sites are included in the set describing the state. For example, for a species $A$ 
with binding sites $\mathsf{sites}(\mathcal{M},A) =  \{a,b\}$, the set 
$\wp( \mathsf{sites}(\mathcal{M},A) ) = \{ \{ \}, \{ a \}, \{ b \}, \{ a,b\} \}$ is the set of all its states. 
Then $\{a\}$  is the state where site 
$a$ on $A$ is bound and site $b$ on $A$ is unbound. We map each sentence
$\langle\, \mathsf{type}, (A,a),\, (B,b), \, \textit{Pos}, \, \textit{Neg},\, r \, \rangle$ 
to a sentence of the form
$$\langle\, \mathsf{type}, (A,a),\, (B,b), \, \mathsf{states}(A), \, \mathsf{states}(B), \, r \, \rangle$$ 
where $\mathsf{states}(A)$ and $\mathsf{states}(B)$ are obtained as follows.  
$$
%
\mathsf{states}(A) = \{\, \mathcal{S} \in \wp( \mathsf{sites}(\mathcal{M}, A))  \; | \;  
( \, (A,x) \in \textit{Pos}   \Rightarrow x \in \mathcal{S} \, )
\land  
( \, x \in \mathcal{S} \Rightarrow  (A,x) \notin \textit{Neg} \, )\, 
\}
$$
This representation allows us to impose another condition on the sentences:
\begin{enumerate}
\item[7.] 
\textbf{There are no overlapping conditions in the sentences.}
For any two sentences of a model $\mathcal{M}$ of the form 
$\langle\, \mathsf{type}_1, (A,a),\, (B,b), \,\textit{Pos}_1 , \, \textit{Neg}_1,  \, r \, \rangle$ and   
$\langle\, \mathsf{type}_2, (A,a),\, (B,b), \,\textit{Pos}_2 , \, \textit{Neg}_2,  \, r \, \rangle$ 
where 
$\mathsf{type}_1  = \mathsf{type}_2$, we obtain  
$\mathsf{states}(A)_1$  and $\mathsf{states}(B)_1,$ for the first  and 
$\mathsf{states}(A)_2$ and  $\mathsf{states}(B)_2,$ for the second sentence. 
Then we have that

$\!\!\!\!\!$-- if
$\mathsf{states}(A)_1 = \mathsf{states}(A)_2$ then it must be that 
$\mathsf{states}(B)_1 \, \cap \;  \mathsf{states}(B)_2 = \emptyset$; 

$\!\!\!\!\!$-- if 
$\mathsf{states}(B)_1 = \mathsf{states}(B)_2$ then it must be that 
$\mathsf{states}(A)_1 \, \cap\;   \mathsf{states}(A)_2 = \emptyset$; 

$\!\!\!\!\!$-- if
$\mathsf{states}(A)_1 \neq \mathsf{states}(A)_2$ and 
$\mathsf{states}(B)_1 \neq  \mathsf{states}(B)_2$ then it must be that   \\
$\mathsf{states}(A)_1\,  \cap \; \mathsf{states}(A)_2 = \emptyset$
and  
$\mathsf{states}(B)_1 \, \cap \;  \mathsf{states}(B)_2 = \emptyset$.
\end{enumerate}

\begin{example}
Consider the models $\mathcal{M}_1$. 
$$
\begin{array}{l}
\mathcal{M}_1 =   \{\,        \langle\, \textit{association}, (A,a),\, (B,b), \, \{ (B,f) \},\, \{ (C,c),\, (A,a), (B,f)\}, \, 1.0 \, \rangle,\\
\quad \quad   \quad \;\:\:   \langle \, \textit{dissociation}, (A,a),\, (B,b), \, \{(B,b)\},\,  \{\}, \,1.0 \, \rangle,\\
\quad \quad   \quad \;\:\:   \langle \, \textit{transformation}, A,\, B, \, \{ \},\,  \{\}, \,1.0 \, \rangle,\\
\quad \quad   \quad \;\:\:   \langle \, \textit{association}, (D,d),\, (E,e), \,\{\},\, \{(D,d), \, (E,e)\}, \,2.0 \, \rangle,\\
\quad \quad   \quad \;\:\:   \langle \, \textit{association}, (D,d),\, (E,e),  \, \{\},\,  \{(D,d),\, (E,e)\}, \, 4.0 \, \rangle \, \}
\end{array}
$$
This model does not fulfill any of the conditions above:  in the first sentence, (1.) $C \notin \{ A,B\}$;
(2.) $(B,f) \in \textit{Pos}$ and  $(B,f) \in \textit{Neg}$;
(3.)  $f \notin \{ b\} $;
(4.) $(B,b) \notin  \{ (C,c),\, (A,a), (B,f)\} $.
In the second sentence,  (5.) $(A,a) \notin  \{ (B,b)\} $.
In the third sentence,  (6.) $\{  \} \neq  \{ (A,a) \} $. 
(7.)  In the fourth and fifth sentences,  $\mathsf{states}(D)_1 = \{ \{ \} \} = \mathsf{states}(D)_2$ and  
 $\mathsf{states}(E)_1 = \{ \{ \} \} = \mathsf{states}(E)_2$.
\end{example}

\begin{example} \label{example:model:2}
The model $\mathcal{M}_2$ fulfills all the conditions above. 
$$
\begin{array}{l}
\mathcal{M}_2 =   \{\,        \langle\, \textit{association}, (A,a_1),\, (B,b), \, \{  \},\, \{ (A,a_1), (B,b)\}, \, 1.0 \, \rangle,\\
\quad \quad   \quad \;\:\:   \langle\, \textit{association}, (A,a_2),\, (C,c), \, \{  \},\, \{ (A,a_2), (C,c)\}, \, 1.0 \, \rangle,\\
\quad \quad   \quad \;\:\:   \langle \, \textit{dissociation}, (A,a_1),\, (B,b), \,\{ (A,a_1), \, (A,a_2), \, (B,b)\},\, \{\} \}, \,2.0 \, \rangle,\\
\quad \quad   \quad \;\:\:   \langle \, \textit{dissociation}, (A,a_1),\, (B,b), \,\{ (A,a_1), \, (B,b)\},\, \{  (A,a_2) \} \}, \,4.0 \, \rangle \, \}
\end{array}
$$
\end{example}


\section{Translation into Stochastic $\pi$ calculus}

We use the representation of the states of species as sets of their sites to map models to stochastic 
$\pi$ calculus specifications. For this purpose, we first map each model to a \emph{compile map}.
Let us first recall some of the definitions of stochastic $\pi$ calculus, implemented in \textsf{SPiM}, 
as they can be found in \cite{PL07}.


\subsection{Stochastic $\pi$ calculus}

\begin{definition}
Syntax of stochastic $\pi$ calculus. Below fn$(P)$ denotes the set of names that are free in $P$.
$$
\begin{array}{ll}
\begin{array}{rll}
E  ::=  &  \quad \emptyset     & \quad\textrm{Empty}\\[2pt]
             | &  \quad E, X(\tilde m)  = P & \quad\textrm{Definition}\\[2pt]
             | &  & \quad \textrm{fn}(P) \subseteq \tilde m \\[2pt]
             \\
\end{array}
\quad & \quad 
\begin{array}{rll}
P,Q  ::=  &  \quad \Sigma     & \quad\textrm{Summation}\\[2pt]
             | &  \quad X(\tilde n) & \quad\textrm{Instance}\\[2pt]
             | &  \quad  P\, | \,  Q & \quad\textrm{Parallel}\\[2pt]
             | &  \quad \verb+new+ \, x \; P & \quad\textrm{Restriction}
\end{array}
\\
\\
\begin{array}{rll}
\Sigma  ::=  &  \quad \mathbf{0}     & \quad\textrm{Null}\\[2pt]
                  |  &  \quad \pi \verb+;+  P +  \Sigma& \quad\textrm{Action}\\[2pt]
             \\
\end{array}
\quad &\quad 
\begin{array}{rll}
\pi  ::=  &  \quad ?x(\tilde m)     & \quad\textrm{Input}\\[2pt]
             | &  \quad !x (\tilde n) & \quad\textrm{Output}\\[2pt]
             | &  \quad  \tau_r & \quad\textrm{Delay}
\end{array}
\end{array}
$$
Expressions above are considered equivalent up to the least congruence relation 
given by the equivalence relation $\equiv$ defined as follows.
$$
\begin{array}{ll}
\begin{array}{rll}
P \; | \; \mathbf{0} & \equiv &  P \\[2pt]
P \; | \; Q & \equiv &  Q \; | \; P \\[2pt]
P \; | \; (Q \; | \; R )& \equiv &  ( P \; | \; Q  ) \; | \; R  \\[2pt]
X(\tilde n)  & \equiv & P\substitute{\tilde n}{\tilde m} \textrm{ if } X(\tilde m) = P 
\end{array}
\quad & \quad 
\begin{array}{rll}
\verb+new+\, x \; \mathbf{0} & \equiv & \mathbf{0}\\[2pt]
\verb+new+\, x \; \verb+new+\, y\;  P  & \equiv & \verb+new+\, y \; \verb+new+\, x\;  P \\[2pt]
\verb+new+\, x \; (  P\; | \; Q ) & \equiv & P \; | \;  \verb+new+\, x \; Q \textrm{ if } x \notin \textrm{fn}(P)\\
\\
\end{array}
\end{array}
$$
\end{definition}


 \subsection{Compile Maps}
 \label{subsection:compile:maps}
 
We map  models into \emph{compile maps}, denoted with $\mathcal{C}$. A compile map is a set 
of expressions that we call \emph{process descriptions} for each species  $A \in \mathsf{species}(\mathcal{M})$.  For a model $\mathcal{M}$,
the process description of species $A \in \mathsf{species}(\mathcal{M})$,  denoted with $\mathsf{P}(A)$, is the pair 
$\langle A, \,\mathsf{actions}(A) \rangle$.
Here, $\mathsf{actions}(A)$ is the set collecting   $\mathsf{actions}(A, \mathcal{S})$
for every $\mathcal{S} \in \wp(\mathsf{sites}(\mathcal{M},A) )$.
$$
\mathsf{actions}(A, \mathcal{S}) 
= 
\langle \mathcal{S},\mathsf{assoc}(A,\mathcal{S}) , \mathsf{dissoc}(A,\mathcal{S}),\mathsf{transform}(A,\mathcal{S}) \rangle  
$$
We define 
$\mathsf{assoc}(A,\mathcal{S})$ as the set of $\mathsf{assoc}(A,\mathcal{S},a)$ 
for every $a \in \mathsf{sites}(\mathcal{M},A)$.
$$
\mathsf{assoc}(A,\mathcal{S},a) = 
\langle 
a, \mathsf{assocPartners}{(A,\mathcal{S},a)}
\rangle
$$
where $\mathsf{assocPartners}(A,\mathcal{S},a)$ is the set 
$$
\begin{array}{l}
\{  \langle B,b, \mathsf{states}(B), r \rangle \; |\;   \\[2pt]
\qquad \qquad \quad 
(\, \langle\, \textit{association}, \, (A,a),\, (B,b),\, \textit{Pos}, \, \textit{Neg},\, r \, \rangle \in \mathcal{M} \; \land \; \mathcal{S} \in \mathsf{states}(A) \, )\\[2pt]
\qquad \quad \quad \lor \;
(\, \langle\, \textit{association}, \, (B,b),\, (A,a),\, \textit{Pos}, \, \textit{Neg},\, r \, \rangle \in \mathcal{M} \; \land \; \mathcal{S} \in \mathsf{states}(A) \, )\, \} \, .
\end{array}
$$
We define 
$\mathsf{dissoc}(A,\mathcal{S})$, similarly,  as the set of $\mathsf{dissoc}(A,\mathcal{S},a)$ 
for every $a \in \mathsf{sites}(\mathcal{M},A)$.
$$
\mathsf{dissoc}(A,\mathcal{S},a) = 
\langle 
a, \mathsf{dissocPartners}{(A,\mathcal{S},a)}
\rangle
$$
where $\mathsf{dissocPartners}(A,\mathcal{S},a)$ is the set 
$$
\begin{array}{l}
\{  \langle B,b, \mathsf{states}(B), r \rangle \; |\;   \\[2pt]
\qquad \qquad \quad 
(\, \langle\, \textit{dissociation}, \, (A,a),\, (B,b),\, \textit{Pos}, \, \textit{Neg},\, r \, \rangle \in \mathcal{M} \; \land \; \mathcal{S} \in \mathsf{states}(A) \, )\\[2pt]
\qquad \quad \quad \lor \;
(\, \langle\, \textit{dissociation}, \, (B,b),\, (A,a),\, \textit{Pos}, \, \textit{Neg},\, r \, \rangle \in \mathcal{M} \; \land \; \mathcal{S} \in \mathsf{states}(A) \, )\, \} \, .
\end{array}
$$
If $\mathcal{S} = \emptyset$, the set  $\mathsf{transform}(A,\mathcal{S}) $ is defined as 
$$
\{  \langle B, r \rangle \; |\;   
(\, \langle\, \textit{transformation}, \, A,\, B,\, \textit{Pos}, \, \textit{Neg}, \, r \, \rangle \in \mathcal{M} \, \} \, .
$$
Otherwise, it is $\emptyset$.

\begin{example}
Consider the model $\mathcal{M}_2$ in Example \ref{example:model:2}. 
We have that the compile map $\mathcal{C}_2$ for this model is as follows.
$$
\begin{array}{ll}
\{ \; \langle\, A, \, \{&\!\!\!\!\! \langle \,  \{ \} ,\,  \{ (a_1, \{ (B,b, \{\{\}\}, 1.0 )\}), \, 
                                                                              (a_2, \{ (C,c, \{\{\}\}, 1.0 ) \},\, \{ \},\, \{\}  \,\rangle\, , \\[2pt]
                               & \!\!\!\!\! \langle \,  \{ a_1 \} ,\,  \{  (a_2, \{ (C,c, \{\{\}\}, 1.0 )  \},\,
                                                                                                                                        \{ (B,b, \{\{ b\}\}, 4.0 )\},\, \{\}  \,\rangle\, , \\[2pt]
                               & \!\!\!\!\! \langle \,  \{ a_2 \} ,\,  \{  (a_1, \{ (B,b, \{\{\}\}, 1.0 ) \},\, \{  \},\, \{\}  \,\rangle\, , \\[2pt]
                               & \!\!\!\!\! \langle \,  \{ a_1,\, a_2 \} ,\,  \{ \},\, \{ (B,b, \{\{ b\}\}, 2.0 ) \},\, \{\}  \,\rangle \; \} \, \rangle\, ,\\[3pt]
\;  \;\; \langle\, B, \, \{ & \!\!\!\!\! \langle \,  \{ \} ,\,  \{  (b, \{ (A,a_1, \{\{\},\, \{ a_2\}\}, 1.0 ) \},\, \{  \},\, \{\}  \,\rangle\, , \\[2pt]
                                   & \!\!\!\!\! \langle \,  \{ b \} ,\,  \{ \},\,
                                       \{ (b, \{ (A,a_1, \{\{ a_1\}\}, 4.0 ), \, (A,a_1, \{\{ a_1,\, a_2\}\}, 2.0 )  \},\, \{\}  \,\rangle\; \} \, \rangle \, , \\[3pt]
\;  \; \; \langle\, C, \, \{  & \!\!\!\!\! \langle \,  \{ \} ,\,  \{ (a_1, \{ (A, a_2,  \{\{\}, \, \{ a_1\}\}, 1.0 )\ \},\, \{  \},\, \{\}  \,\rangle\, , \\[2pt]
                                   & \!\!\!\!\! \langle \,  \{ c \} ,\,  \{ \},\, \{ \},\, \{\}  \,\rangle\; \} \, \rangle \,  \} 
\end{array}
$$
%
%
%
\end{example}

\subsection{From Compile Maps to Stochastic $\pi$ calculus}
\label{subsection:translation} 

We construct a $\pi$ calculus specification from the  
compile map $\mathcal{C}$ of a model $\mathcal{M}$.
For each species $A \in \mathsf{species}(\mathcal{M})$, we map 
the process description $\mathsf{P}(A)$ to a process specification
in stochastic $\pi$ calculus.
Let  
$$
\mathsf{P}(A) = \langle A, \{ \, \mathsf{actions}(A, \mathcal{S}_1), \ldots,  \mathsf{actions}(A, \mathcal{S}_n) \, \} \rangle
$$
where $\wp(\mathsf{sites}(\mathcal{M},A)) = \{ \mathcal{S}_1,\ldots,\mathcal{S}_n \}$, that is, the powerset of set of sites of $A$.
Thus,  there are $n$ process specifications 
for the species $A$, some of which may be empty. 
Each process specification for each state $\mathcal{S}$ of $A$ is of the following syntactic form.
$$
\begin{array}{lll}
 \textit{process declaration} &
\: \textrm{``=  (''} & \textit{ local channel declarations}\\[2pt]
                                        & \;\;\;  & \textit{ association specifications}\\[2pt]  
                                        &\;  \; \;  \textrm{``+''} & \textit{ dissociation specifications}\\[2pt] 
                                       & \; \; \;\textrm{``+''} & \textit{ transformation specifications} \;\;  \textrm{``)''} \; 
\end{array}                                     
$$

The idea here is that each set of sites of a species $A$ denotes the state where the sites in the set are bound. 
Thus the powerset of the set of sites of a species denotes the set of all its states. 
Now, let us obtain the process expression for each state $\mathcal{S}_i$
with respect to $\mathsf{actions}(A, \mathcal{S}_i)$ where $1 \leq i \leq n$. 
Let us consider  $\mathcal{S}_i = \{a_1,\ldots, a_k\}$ of $A$ with  
$$
\mathsf{actions}(A, \mathcal{S}_i)= \langle \mathcal{S}_i \, ,
\mathsf{assoc}(A,\mathcal{S}_i) , \mathsf{dissoc}(A,\mathcal{S}_i),
\mathsf{transform}(A,\mathcal{S}_i)  \rangle \, .
$$

\subsubsection*{Process declaration}
The expression for  \textit{process declaration} is a process name with its list of parameters.
It is delivered by the dissociation sentences in $\mathcal{M}$ and 
  $\mathcal{S}_i = \{a_1,\ldots, a_k\}$. 
For every $a_j \in \mathcal{S}_i$,
consider the set 
$$
\begin{array}{l}
\!\!\!\!\! \mathcal{R}(A,a_j) = \\[2pt]
\!\!\!\!\!  \{  (a_j ,\, (r/2))\; | \,  \langle  \textit{dissociation}, (A,a_j), (B,b), \, \textit{Pos},\, \textit{Neg},\,  r  \rangle  \in  \,  \mathcal{M}  \, \}
 \; \cup\\[2pt]
 \!\!\!\!\!  \,\quad \{  (a_j ,\, (r/2))\; | \, \langle  \textit{dissociation}, (B,b),  (A,a_j), \, \textit{Pos},\, \textit{Neg},\,  r  \rangle \, \in \, \mathcal{M} \, \} \; \cup\\[2pt]
 \!\!\!\!\! 
 \qquad  \,\,\quad  \{  (a_j ,\, 1.0)\; | \, \langle  \textit{dissociation}, (B,b),  (A,a_j),  \, \textit{Pos},\, \textit{Neg},\,  r  \rangle \, \notin \, \mathcal{M} \; \land \; \\[2pt]
 \qquad \qquad \quad \qquad \; \;  \quad   \langle  \textit{dissociation},  (A,a_j), (B,b), \, \textit{Pos},\, \textit{Neg},\,  r  \rangle \,  \notin \, \mathcal{M}  \}  \, .      
\end{array}
$$
We associate each element of the set 
$\mathcal{R}(A, a_j)$ a unique label $s \in \mathbb{N}^+$ and obtain $\mathcal{R}'(A, a_j)$. Then if 
$
\mathcal{R}'(A, a_j) = \{  (a_j ,\, r_1,\, 1), \ldots, (a_j ,\, r_\ell, \, \ell) \, \}
$
 we write the process declaration 
for $A$ at state $\mathcal{S}_i  = \{a_1,\ldots, a_k\}$  as follows.
$$
A_i( {a_1}{1}, \ldots ,{a_1}{\ell_1},\ldots\ldots,{a_k}{1}, \ldots ,{a_k}{\ell_k})
$$

\begin{example}
For the state $\mathcal{S}_2 = \{ a_1 \}$ of species $A$ of Example \ref{example:model:2}, 
we have the process declaration below, since  we have that $\,\mathcal{R}'(A, a_1) = \{ (a_1,2.0,1),(a_1,1.0,2)\}$.  
$$
A_2({a_1}{1},{a_1}{2})
$$

\end{example}

\subsubsection*{Local channel declarations}
These expressions are delivered by the dissociation 
sentences in $\mathcal{M}$ and the $\mathsf{assoc}(A,\mathcal{S}_i)$.  That is, for every 
$$ 
\mathsf{assoc}(A,\mathcal{S}_i, a_j) = \langle\,  a_j , \, \mathsf{assocPartners}(A,\mathcal{S}_i,a_j) \, \rangle \in \mathsf{assoc}(A,\mathcal{S}_i)\, ,
$$
and for every 
$
\langle \, B, b, \mathsf{states}(B),\, r \, \rangle \in \mathsf{assocPartners}(A,\mathcal{S}_i,a_j)
$
consider the set 
$$
\begin{array}{l}
\!\!\!\!\! \mathcal{U}(A, a_j,B,b) = \\[2pt]
\!\!\!\!\!  \{  (a_j ,\, (r/2))\; | \,  \langle  \textit{dissociation}, (A,a_j), (B,b), \, \textit{Pos},\, \textit{Neg},\,  r  \rangle  \in  \,  \mathcal{M} \, \land \,
                 a_j \prec b \, \}
 \; \cup\\[2pt]
 \!\!\!\!\!  \,\quad \{  (a_j ,\, (r/2))\; | \, \langle  \textit{dissociation}, (B,b),  (A,a_j), \, \textit{Pos},\, \textit{Neg},\,  r  \rangle \, \in \, \mathcal{M} \, 
        \land \,         a_j \prec b \, \} \; \cup\\[2pt]
 \!\!\!\!\! 
  \,\,\quad  \quad \{  (a_j ,\, 1.0)\; | \, \langle  \textit{dissociation}, (B,b),  (A,a_j),  \, \textit{Pos},\, \textit{Neg},\,  r  \rangle \, \notin \, \mathcal{M} \; \land \; \\[2pt]
 \quad  \quad \qquad \qquad \; \;  \quad   \langle  \textit{dissociation}, (A,a_j),  (B,b), \, \textit{Pos},\, \textit{Neg},\,  r  \rangle \,  \notin \, \mathcal{M} 
    \,  \land \,         a_j \prec b \,  \}        
\end{array}
$$
where  $\prec$ denotes a lexicographic order on  sites. We associate each element of the set 
$\mathcal{U}(A, a_j,B,b)$ a unique label $s \in \mathbb{N}^+$ to obtain $\mathcal{U}'(A, a_j,B,b)$. Then if 
$$
\mathcal{U}'(A,a_j,B,b) = \{  (a_j ,\, r_1,\, 1), \ldots, (a_j ,\, r_\ell,\, \ell) \}
$$
then we write the channel declarations 
for $\mathsf{assoc}(A,\mathcal{S}_i, a_j)$  as follows.
$$
 \verb+new + {a_j}{1}\verb+@+r_1 \quad \ldots \quad \verb+new + {a_j}{\ell}\verb+@+r_{\ell}
$$

\begin{example}
For the state $\mathcal{S}_2 = \{ a_1 \}$ of species $A$ of Example \ref{example:model:2}, 
we have the channel declarations below, since  we have that  $\mathcal{U}'(A,  a_2,B,b) = \{ (a_2,1.0,1) \}$.  
$$
\verb+new + {a_2}{1}\verb+@+1.0
$$
\end{example}

\subsubsection*{Association specifications}
The expression for  \textit{association specifications} for species $A$ at state  $\mathsf{assoc}(A,\mathcal{S}_i)$
is delivered by  $\mathsf{assoc}(A,\mathcal{S}_i)$. For every 
$$ 
\langle\,  a_j , \, \mathsf{assocPartners}(A,\mathcal{S}_i,a_j) \, \rangle \in \mathsf{assoc}(A,\mathcal{S}_i),
$$
and for every 
$
\langle \, B, b, \mathsf{states}(B),\, r \, \rangle \in \mathsf{assocPartners}(A,\mathcal{S}_i,a_j)
$
consider the set 
$$
\begin{array}{l}
\!\!\!\!\! \mathcal{B}(A, a_j,B,b) = \\[2pt]
\quad \quad \!\!\!\!\!  \{  (\verb+!+a_jb, \, r)\; | \,  \langle  (B,b), \, \mathsf{states}(B),\,   r  \rangle  \in
 \mathsf{assocPartners}(A,\mathcal{S}_i,a_j) \, \land \, a_j \prec b \, \}
  \; \, \cup\\[2pt]
\quad \quad \quad \!\!\!\!\!  \{  (\verb+?+ba_j ,\, r)\; | \,  \langle  (B,b), \, \mathsf{states}(B),\,   r  \rangle  \in 
\mathsf{assocPartners}(A,\mathcal{S}_i,a_j) \, \land \, b \prec a_j \, \} \, .
\end{array}
$$
We associate each element of the set 
$\mathcal{B}(A, a_j,B,b)$ a unique label $s \in \mathbb{N}^+$ and obtain $\mathcal{B}'(A, a_j,B,b)$.  
Association of  site $a_j$ on $A$ results in the state $\mathcal{S}_{i'} =\mathcal{S}_i \cup \{ a_j \}$. 
For each element of  $(\verb+!+a_jb , r_s,s) \in  \mathcal{B}'(A, a_j,B,b)$
we write the following,  composed by \verb-+-. 
$$
 \verb+!+a_jbs(a_j1,\ldots,a_j\ell)\verb+;+  \textit{continuation}  \;
$$
The association channel names, such as  $a_jbs$ here, 
are also declared as \emph{global channel declarations},
preceding all the process declarations. 
The \textit{continuation} 
is written  for A in  $\mathcal{S}_{i'}$ as for \emph{process declarations} above,
however we write \verb+nil+ for the channel names for those associations of site $a_j$ on $A$
with some site $b' \neq b$. Here, \verb+nil+ is the nil-dissociation channel with rate $0$.    
We obtain $a_j1,\ldots,a_j\ell$ from the set $\mathcal{U}(A,a_j,B,b)$ as in 
\emph{channel declarations}. 

\begin{example}
For the state $\mathcal{S}_2 = \{ a_1 \}$ of species $A$ of Example \ref{example:model:2}, 
we have the following association specifications.
$$
 !a2c1(a_2)\verb+;+ A3(a_11,a_12,a_2) 
$$
\end{example}

\subsubsection*{Dissociation specifications}
The expression for  \textit{dissociation specifications} for species $A$ at state  $\mathsf{assoc}(A,\mathcal{S}_i)$
is delivered by  $\mathsf{dissoc}(A,\mathcal{S}_i)$. For every 
$$ 
\langle\,  a_j , \, \mathsf{dissocPartners}(A,\mathcal{S}_i,a_j) \, \rangle \in \mathsf{dissoc}(A,\mathcal{S}_i),
$$
and for every 
$
\langle \, B, b, \mathsf{states}(B),\, r \, \rangle \in \mathsf{dissocPartners}(A,\mathcal{S}_i,a_j)
$
consider the set 
$$
\begin{array}{l}
\!\!\!\!\! \mathcal{G}(A, a_j,B,b) = \\[2pt]
\quad \quad \!\!\!\!\!  \{  (\verb+!+a_j, \, r)\; | \,  \langle  (B,b), \, \mathsf{states}(B),\,   r  \rangle  \in
 \mathsf{dissocPartners}(A,\mathcal{S}_i,a_j) \, \land \, a_j \prec b \, \}
 \; \, \cup\\[2pt]
\quad \quad \quad \!\!\!\!\!  \{  (\verb+?+b ,\, r)\; | \,  \langle  (B,b), \, \mathsf{states}(B),\,   r  \rangle  \in 
\mathsf{dissocPartners}(A,\mathcal{S}_i,a_j) \, \land \, b \prec a_j \, \} \, .
\end{array}
$$
We associate each element of the set 
$\mathcal{G}(A, a_j,B,b)$ a unique label $s \in \mathbb{N}^+$ and obtain $\mathcal{G}'(A, a_j,B,b)$.  
Dissociation of  $a_j$ on $A$ results in the state $\mathcal{S}_{i'} =\mathcal{S}_i \setminus \{ a_j \}$. 
For each $(\verb+!+a_j, r_s,s) \in  \mathcal{G}'(A, a_j,B,b)$
we write the following,  composed by ``\verb-+-''.
$$
 \verb+!+a_js\verb+;+  \textit{continuation}  \;\verb- + -  \: \verb+?+a_js\verb+;+  \textit{continuation} 
$$
The \textit{continuation} 
is written  for A in  $\mathcal{S}_{i'}$ as for \emph{process declarations} above.

\begin{example}
For the state $\mathcal{S}_2 = \{ a_1 \}$ of species $A$ of Example \ref{example:model:2}, 
we have the following dissociation specifications.
$$
 \verb+!+a_11\verb+;+ A_1() \verb- + - \verb+?+a_11; A_1() 
$$
\end{example}

\subsubsection*{Transformation specifications}
The expression for  \textit{transformation specifications} for species $A$ is given only 
if the  state  $\mathcal{S} = \{\}$. In that case, for  
 $\mathsf{transfrom}(A,\{\}) = \{(B_1,r_1), \ldots, (B_k, r_k) \}$ we write
 $$
\verb+delay@+r_1\verb+;+ B_1()\verb- + - \ldots \verb- + - \verb+delay@+r_k\verb+;+ B_k()
 $$

\section{Syntax of the Language}
\label{section:language}

The syntax of the language is defined in BNF 
notation, where optional elements are enclosed in braces as \{Optional\}. 
A model (description) consists of sentences of the following form.\\


\begin{tabular}{lrl}
\textit{Model}  & \grameq & \textit{Sentence}$_1$ \ldots \space \textit{Sentence}$_m \qquad m \geq 1$\\[4pt]
\textit{Sentence}  & \grameq   & \textit{Association}             \\
                   & $|$       & \textit{Dissociation}        \\
                   & $|$       & \textit{Transformation}      \\
                   & $|$       & \textit{Decay}               \\
                   & $|$       & \textit{Phosphorylation}     \\
                   & $|$       & \textit{Dephosphorylation}   \\[4pt]
\textit{Association}   & \grameq   & \textit{Site} \texttt{on} \textit{Species} \texttt{associates}
                                 \textit{Site} \texttt{on} \textit{Species}\\ 
& &    $\quad \qquad \qquad$  \{\texttt{with rate} \textit{Float}\} 
\{\texttt{if} \textit{Conditions}\}\\[4pt]   
\textit{Dissociation} & \grameq   & \textit{Site} \texttt{on} \textit{Species} \texttt{dissociates}
                                 \textit{Site} \texttt{on} \textit{Species}\\ 
& &    $\quad \qquad \qquad$  \{\texttt{with rate} \textit{Float}\} 
\{\texttt{if} \textit{Conditions}\}\\[4pt]   
\textit{Phosphorylation}   & \grameq   & \textit{Site} \texttt{on} \textit{Species} \texttt{gets phosphorylated} \\
& &    $\quad \qquad \qquad$  
\{\texttt{with rate} \textit{Float}\} \{\texttt{if}  \textit{Conditions}\}\\[4pt]
\textit{Dephosphorylation}   & \grameq   & \textit{Site} \texttt{on} \textit{Species} \texttt{gets dephosphorylated} \\
& &    $\quad \qquad \qquad$  
\{\texttt{with rate} \textit{Float}\} \{\texttt{if} \textit{Conditions}\} 
\end{tabular}
\\
\begin{tabular}{lrl} 
\textit{Transformation} & \grameq   & \textit{Species} \texttt{becomes} \textit{Species} 
                             \{\texttt{with rate} \textit{Float}\} 
\\[4pt]   
\textit{Decay} & \grameq   & \textit{Species} \texttt{decays}
                             \{\texttt{with rate} \textit{Float}\} 
\\[4pt]
\textit{Conditions}   & \grameq   & \textit{Condition}\\
& $|$ &   \textit{Condition} \texttt{and} \textit{Conditions}\\[4pt]
\textit{Condition}   & \grameq & \textit{Site} \texttt{on} \textit{Species} \texttt{is bound}\\
                & $|$     & \textit{Site} \texttt{on} \textit{Species} \texttt{is unbound}\\[4pt]
\textit{Site}   & \grameq   & \textit{String}\\[4pt]
\textit{Species}   & \grameq   & \textit{String}

\end{tabular}
\\[4pt]

In our implementation of the translation algorithm, 
each sentence of a model given in this syntax  
is mapped by a lexer and parser 
to a data structure of the form given in Section \ref{section:data:structure} in the obvious way.  
Phosphorylation sentences are treated as association sentences 
where the second species is  by default \verb+Phosph+ with the binding site \verb+phosph+. The 
dephosphorylation sentences are mapped similarly to dissociation sentences. If not given,
a default rate ($1.0$) is assigned to sentences.

\subsection{A Model of Fc$\gamma$ Receptor-mediated Phagocytosis}

We demonstrate the use of the language on a model of Fc$\gamma$ receptor (Fc$\gamma$R) phosphorylation 
during phagocytosis, where the binding of complexed immunoglobulins G (IgG) to Fc$\gamma$R triggers a 
signalling cascade that leads to actin-driven particle engulfment \cite{GR02,SH04,CHDLC06}. When a small particle is coated 
(opsonised) with IgG, the Fc regions of the IgG molecules can bind to Fc$\gamma$Rs in the plasma membrane 
and initiate a phagocytic response: a signalling cascade then drives the remodelling of the actin cytoskeleton 
close to the membrane. This results in cup-shaped folds of plasma membrane that extend outwards around the
 internalised particle and eventually close into a plasma membrane-derived phagosome.

\begin{figure}[t]
$$
\includegraphics[width=150mm]{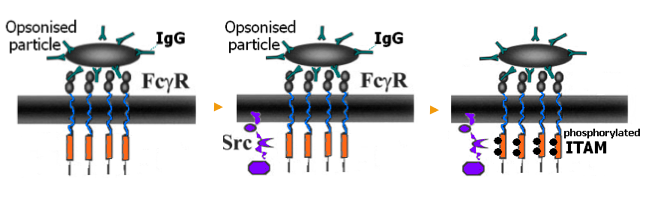}
$$
\caption{A simple model of the phosphorylation of the ITAM domain on the Fc$\gamma$ receptor during phagocytosis. 
Adapted from \cite{GR02}.
\label{figure:fcgamma}
 }
\end{figure}

Fc$\gamma$R contains within its cytoplasmic tail an immunoreceptor tyrosine-based activation motif (ITAM). 
The association of Fc$\gamma$R 
with an IgG induces the phosphorylation of two tyrosine residues within the ITAM domain by Src-family kinases. 
The  phosphorylated ITAM domain then recruits Syk kinase, which propagates the signal further to downstream 
effectors (see Figure \ref{figure:fcgamma}). 
In our  language, we can describe the initial phases of this cascade as follows: 

{\footnotesize
\begin{verse} 
\texttt{site f on FcR associates site i on IgG with rate 2.0} \\[1pt]
\texttt{site y on FcR gets phosphorylated if site f on FcR is bound}\\[1pt] 
\texttt{site z on FcR gets phosphorylated if site f on FcR is bound}  
\end{verse}
}
 The first sentence above describes the binding of Fc$\gamma$R to IgG. 
 The second and third sentences describe the phosphorylation  of 
the two tyrosine residues on ITAM (association of a \verb+Phosph0()+ molecule). 
This is automatically translated by our tool into 
the SPiM program given in Appendix A. 
We can then run stochastic simulations on the model given by these sentences.

By using this language and our translation tool, we can build models of different size and complexity, 
and modify and extend  these models with respect to the knowledge in hand on the different sites and
 interaction capabilities of the Fc$\gamma$R, as well as  other biological systems. For example, the 
 model above abstracts away from the role played by the Src kinases in the  phosphorylation of the 
 Fc$\gamma$R as depicted in Figure \ref{figure:fcgamma}. 
 The sentences above can be easily modified and extended to capture this aspect 
 in the model as follows: here, 
 the shaded part demonstrates the modifications with respect to the model given above.

{\footnotesize
\begin{verse}
\texttt{site f on FcR associates site i on IgG with rate 2.0}\\[1pt]
\texttt{site y on FcR gets phosphorylated if site \rdx{s} on FcR is bound}\\[1pt]
\texttt{site z on FcR gets phosphorylated if site \rdx{s} on FcR is bound}\\[1pt]
\texttt{site s on FcR associates site sr on Src if site f on FcR is bound}\\[1pt]
\texttt{site s on FcR dissociates site sr on Src}\\[1pt]
\end{verse}
}

The $\mathsf{SPiM}$ program resulting from automated translation of this model is given in Appendix B. 
It is important to note that, because FcR has 4 binding sites in the model above, in the $\mathsf{SPiM}$
code resulting from the translation, there are 16 species for FcR, denoting its different possible states. 
However, in the code given in Appendix A, there are 8 species for FcR denoting its states that result 
from its 3 binding sites in that model. 

\section{Discussion}
We have introduced a natural language interface for building stochastic $\pi$ calculus models 
of biological systems.  The $\kappa$-calculus  \cite{DFFHK07,DFFHK08,DFFK08} 
and the work on Beta-binders in \cite{GHP07} 
have been source of inspiration for this language.

 In \cite{GHP07}, Guerriero et al. give a narrative style 
interface for the process algebra Beta-binders for a rich biological language. 
In our language, we build complex events such as phosphorylation 
and dephosphorylation of sites as instances of basic primitives  of association, 
dissociation and transformation.  We give a functional translation algorithm for 
our translation into stochastic $\pi$ calculus. 
The conditions that we impose on the models
are automatically verified in the implementation of our tool.
These  conditions  should be 
instrumental for `debugging' purposes while building increasingly large models.

The implicit semantics of our language,  which is 
implemented in the translation algorithm into $\pi$ calculus, 
can be seen as a translation of a fragment of the $\kappa$ 
calculus into $\pi$ calculus.
Another approach similar to the one in this paper is the work by  Laneve et al. 
in \cite{PLZ09},  where the authors give an encoding of nano-$\kappa$-calculus 
in $\mathsf{SPiM}$.  In comparison with our algorithm,  
the encoding in \cite{PLZ09} covers a larger part
of nano-$\kappa$ by using  
the $\mathsf{SPiM}$ language as a programming language for
implementing a notion of term rewriting, where there is an explicit function for
matching.  The algorithm  gives the different states of a species in the $\mathsf{SPiM}$ encoding 
with respect to the parameters of that species as in $\kappa$-calculus.

Topics of future work  include
an exploration of the expressive power of the association, dissociation and transformation 
primitives with respect to  Kohn diagram representation \cite{KAKWP06} of biological models.

\bibliographystyle{plain}
\bibliography{ozan.bib}

\begin{thebibliography}{10}

\bibitem{CCGKP08b}
L.~Cardelli, E.~Caron, P.~Gardner, O.~Kahramano{\u g}ullar{\i}, and
  A.~Phillips.
\newblock A process model of actin polymerisation.
\newblock In {\em FBTC'08}, volume 229 of {\em ENTCS}, pages 127--144.
  Elsevier, 2008.

\bibitem{CCGKP08}
L.~Cardelli, E.~Caron, P.~Gardner, O.~Kahramano{\u g}ullar{\i}, and
  A.~Phillips.
\newblock A process model of {Rho GTP}-binding proteins.
\newblock {\em Theoretical Computer Science}, 410/33-34:3166--3185, 2009.

\bibitem{CZ08b}
L.~Cardelli and G.~Zavattaro.
\newblock On the computational power of biochemistry.
\newblock In {\em AB'08}, volume 5147 of {\em LNCS}, pages 65--80. Springer,
  2008.

\bibitem{CHDLC06}
C.~Cougoule, S.~Hoshino, A.~Dart, J.~Lim, and E.~Caron.
\newblock Dissociation of recruitment and activation of the small {G}-protein
  {Rac} during {Fc gamma} receptor-mediated phagocytosis.
\newblock {\em J. Bio. Chem.}, 281:8756--8764, 2006.

\bibitem{DFFHK07}
V.~Danos, J.~Feret, W.~Fontana, R.~Harmer, and J.~Krivine.
\newblock Rule-based modelling of cellular signalling.
\newblock In {\em CONCUR'07}, volume 4703 of {\em LNCS}, pages 17--41.
  Springer, 2007.

\bibitem{DFFHK08}
V.~Danos, J.~Feret, W.~Fontana, R.~Harmer, and J.~Krivine.
\newblock Rule-based modelling, symmetries, refinements.
\newblock In {\em FMSB'08}, volume 5054 of {\em LNCS}, pages 103--122.
  Springer, 2008.

\bibitem{DFFK08}
V.~Danos, J.~Feret, W.~Fontana, and J.~Krivine.
\newblock Abstract interpretation of cellular signalling networks.
\newblock In {\em VMCAI'08}, volume 4905 of {\em LNBI}, pages 83--97. Springer,
  2008.

\bibitem{GR02}
E.~Garcia-Garcia and C.~Rosales.
\newblock Signal transduction during {Fc} receptor-mediated phagocytosis.
\newblock {\em Journal of Leukocyte Biology}, 72:1092--1108, 2002.

\bibitem{HGD08}
M.~Heiner, D.~Gilbert, and R.~Donaldson.
\newblock Petri nets for systems and synthetic biology.
\newblock In {\em SFM'08}, volume 5016 of {\em LNCS}, pages 215--264. Springer,
  2008.

\bibitem{KAKWP06}
K.~W. Kohn, M.~I. Aladjem, S.~Kim, J.~N. Weinstein, and Y.~Pommier.
\newblock Depicting combinatorial complexity with the molecular interaction map
  notation.
\newblock {\em Molecular Systems Biology}, 2:51, 2006.

\bibitem{GHP07}
C.~Priami M.~L.~Guerriero, J. K.~Heath.
\newblock An automated translation from a narrative language for biological
  modelling into process algebra.
\newblock In {\em CMSB'07}, volume 4695 of {\em LNCS}, pages 136--151.
  Springer, 2007.

\bibitem{Mil99}
R.~Milner.
\newblock {\em Communication and Mobile Systems: the {$\pi$}-calculus}.
\newblock Cambridge University Press, 1999.

\bibitem{PL07}
A.~Phillips and L.~Cardelli.
\newblock Efficient, correct simulation of biological processes in stochastic
  {Pi}-calculus.
\newblock In {\em CMSB'07}, volume 4695 of {\em LNBI}. Springer, 2007.

\bibitem{PCC06}
A.~Phillips, L.~Cardelli, and G.~Castagna.
\newblock A graphical representation for biological processes in the stochastic
  pi-calculus.
\newblock In {\em Transactions on Computational Systems Biology {VII}}, volume
  4230 of {\em LNCS}, pages 123--152. Springer, 2006.

\bibitem{PLZ09}
S.~Pradalier, C.~Laneve, and G.~Zavattaro.
\newblock From biochemistry to stochastic processes.
\newblock In {\em QALP'09}, ENTCS. Elsevier, 2009.
\newblock to appear.

\bibitem{PRSS01}
C.~Priami, A.~Regev, E.~Shapiro, and W.~Silverman.
\newblock Application of a stochastic name-passing calculus to representation
  and simulation of molecular processes.
\newblock {\em Information Processing Letters}, 80:25--31, 2001.

\bibitem{RS02}
A.~Regev and E.~Shapiro.
\newblock Cellular abstractions: Cells as computation.
\newblock {\em Nature}, 419:343, 2002.

\bibitem{SH04}
J.~A. Swanson and A.~D. Hoppe.
\newblock The coordination of signaling during {Fc} receptor-mediated
  phagocytosis.
\newblock {\em Journal of Leukocyte Biology}, 76:1093--1103, 2004.

\bibitem{TTKLL07}
A.~Tiwari, C.~Talcott, M.~Knapp, P.~Lincoln, and K.~Laderoute.
\newblock Analyzing pathways using sat-based approaches.
\newblock In Gerhard Goos, Juris Hartmanis, and Jan van Leeuwen, editors, {\em
  Second International Conference, Algebraic Biology 2007}, volume 4545 of {\em
  LNCS}, pages 155--169. Springer, 2007.

\end{thebibliography}

\newpage

\section*{Appendix A}

{\footnotesize
\begin{verse} 
\texttt{site f on FcR associates site i on IgG with rate 2.0} \\[1pt]
\texttt{site y on FcR gets phosphorylated if site f on FcR is bound}\\[1pt] 
\texttt{site z on FcR gets phosphorylated if site f on FcR is bound}  
\end{verse}
}
The $\mathsf{SPiM}$ code resulting from the automated translation of this model.  

\begin{multicols}{2}
{\footnotesize
\begin{verbatim}
directive sample 10.0 
directive plot FcR7(); FcR6(); 
        FcR5();  FcR4(); FcR3(); 
        FcR2(); FcR1();  
        FcR0(); IgG1(); IgG0();  
        Phosph1(); Phosph0() 
 
new fi1@1.0:chan(chan)   
new phosphy2@1.0:chan(chan)   
new phosphz3@1.0:chan(chan)   
new nil@0.0:chan 
 
let FcR0() =  
     ( new f@1.0:chan  
       !fi1(f)*2.0; FcR1(f) ) 
 
 
and FcR1(f:chan) =  
     ( do ?phosphy2(y); FcR4(f,y)  
       or ?phosphz3(z); FcR5(f,z) ) 
 
and FcR2(y:chan) =  
     ( new f@1.0:chan  
       !fi1(f)*2.0; FcR4(f,y) ) 
 
and FcR3(z:chan) =  
     ( new f@1.0:chan  
       !fi1(f)*2.0; FcR5(f,z) ) 
 
and FcR4(f:chan,y:chan) =  
     ( ?phosphz3(z); FcR7(f,y,z) ) 
 
 
and FcR5(f:chan,z:chan) =  
     ( ?phosphy2(y); FcR7(f,y,z) ) 
 
and FcR6(y:chan,z:chan) =  
     ( new f@1.0:chan  
       !fi1(f)*2.0; FcR7(f,y,z) ) 
 
and FcR7(f:chan,y:chan,z:chan) =  
     ()  
 
let IgG0() =  
     ( ?fi1(i); IgG1(i) ) 
 
and IgG1(i:chan) =  
     () 
 
let Phosph0() =  
     ( new phosph@1.0:chan  
       do !phosphy2(phosph)*1.0; 
            Phosph1(phosph)  
       or !phosphz3(phosph)*1.0; 
            Phosph1(phosph) ) 
 
and Phosph1(phosph:chan) =  
     () 
 
 
run 1000 of  FcR0() 
run 1000 of  IgG0() 
run 1000 of  Phosph0() 
\end{verbatim}
}
\end{multicols}


\section{Appendix B}

{\footnotesize
\begin{verse}
\texttt{site f on FcR associates site i on IgG with rate 2.0}\\[1pt]
\texttt{site y on FcR gets phosphorylated if site \rdx{s} on FcR is bound}\\[1pt]
\texttt{site z on FcR gets phosphorylated if site \rdx{s} on FcR is bound}\\[1pt]
\texttt{site s on FcR associates site sr on Src if site f on FcR is bound}\\[1pt]
\texttt{site s on FcR dissociates site sr on Src}\\[1pt]
\end{verse}
}

The $\mathsf{SPiM}$ code resulting from the automated translation of this model.  

\begin{multicols}{2}
{\footnotesize
\begin{verbatim}

directive sample 10.0
directive plot FcR15();
      FcR14(); FcR13(); FcR12(); 
      FcR11(); FcR10();
      FcR9(); FcR8(); FcR7();
      FcR6(); FcR5();
      FcR4(); FcR3(); FcR2(); 
      FcR1(); FcR0();
      IgG1(); IgG0();
      Phosph1(); Phosph0();
      Src1(); Src0()

new fi1@1.0:chan(chan)
new phosphx2@1.0:chan(chan)
new phosphy3@1.0:chan(chan)
new ssr4@1.0:chan(chan)
new nil@0.0:chan


let FcR0() =
 ( new f@1.0:chan
    !fi1(f)*2.0; FcR1(f) )

and FcR1(f:chan) =
 ( new s1@0.50:chan
   !ssr4(s1)*1.0; FcR5(f,s1) )

and FcR2(s1:chan) =
 ( new f@1.0:chan
   do !fi1(f)*2.0; FcR5(f,s1)
   or ?phosphx2(x); FcR8(s1,x)
   or ?phosphy3(y); FcR9(s1,y)
   or !s1; FcR0() or ?s1; FcR0() )

and FcR3(x:chan) =
 ( new f@1.0:chan
   !fi1(f)*2.0; FcR6(f,x) )

and FcR4(y:chan) =
 ( new f@1.0:chan
   !fi1(f)*2.0; FcR7(f,y) )

and FcR5(f:chan,s1:chan) =
 ( do ?phosphx2(x); FcR11(f,s1,x)
   or ?phosphy3(y); FcR12(f,s1,y)
   or !s1; FcR1(f) or ?s1; FcR1(f) )

and FcR6(f:chan,x:chan) =
 ( new s1@0.50:chan
   !ssr4(s1)*1.0; FcR11(f,s1,x) )

and FcR7(f:chan,y:chan) =
 ( new s1@0.50:chan
   !ssr4(s1)*1.0; FcR12(f,s1,y) )

and FcR8(s1:chan,x:chan) =
 ( new f@1.0:chan
   do !fi1(f)*2.0; FcR11(f,s1,x)
   or ?phosphy3(y); FcR14(s1,x,y)
   or !s1; FcR3(x) or ?s1; FcR3(x) )

and FcR9(s1:chan,y:chan) =
 ( new f@1.0:chan
   do !fi1(f)*2.0; FcR12(f,s1,y)
   or ?phosphx2(x); FcR14(s1,x,y)
   or !s1; FcR4(y) or ?s1; FcR4(y) )

and FcR10(x:chan,y:chan) =
 ( new f@1.0:chan
   !fi1(f)*2.0; FcR13(f,x,y) )

and FcR11(f:chan,s1:chan,x:chan) =
 ( do ?phosphy3(y); FcR15(f,s1,x,y)
   or !s1; FcR6(f,x) or ?s1; FcR6(f,x) )

and FcR12(f:chan,s1:chan,y:chan) =
 ( do ?phosphx2(x); FcR15(f,s1,x,y)
   or !s1; FcR7(f,y) or ?s1; FcR7(f,y) )

and FcR13(f:chan,x:chan,y:chan) =
 ( new s1@0.50:chan
   !ssr4(s1)*1.0; FcR15(f,s1,x,y) )

and FcR14(s1:chan,x:chan,y:chan) =
 ( new f@1.0:chan
   do !fi1(f)*2.0; FcR15(f,s1,x,y)
   or !s1; FcR10(x,y) or ?s1; FcR10(x,y) )

and FcR15(f:chan,s1:chan,x:chan,y:chan) =
 ( do !s1; FcR13(f,x,y) or ?s1; FcR13(f,x,y) )


 let IgG0() =
   ( ?fi1(i); IgG1(i) )

 and IgG1(i:chan) =
   ()


 let Phosph0() =
   ( new phosph@1.0:chan
     do !phosphx2(phosph)*1.0; 
     Phosph1(phosph)
     or !phosphy3(phosph)*1.0; 
     Phosph1(phosph) )

 and Phosph1(phosph:chan) =
   ()


 let Src0() =
   ( ?ssr4(sr1); Src1(sr1) )

 and Src1(sr1:chan) =
   ( do !sr1; Src0() or ?sr1; Src0() )


 (* run  1000 of ... *)
\end{verbatim}
}
\end{multicols}


\end{document}